\documentclass[12pt,preprint]{aastex}

\shorttitle{An ultracompact X-ray binary in M\,15}
\shortauthors{Dieball et al.}

\begin{document}

\title{An ultracompact X-ray binary in the globular cluster M\,15
  (NGC\,7078)$^1$}
\footnotetext[1]{Based on observations with the NASA/ESA Hubble Space
  Telescope, obtained at the Space Telescope Science Institute, which
  is operated by the Association of Universities for Research in
  Astronomy, Inc. under NASA contract No. NAS5-26555.} 

\author{A.~Dieball$^2$, C.~Knigge$^2$, D.~R.~Zurek$^3$,
  M.~M.~Shara$^3$, K.~S.~Long$^4$, P.~A.~Charles$^{5,2}$,
  D.~C.~Hannikainen$^6$ and L.~van~Zyl$^{7,8}$}
\footnotetext[2]{School of Physics and Astronomy, University of Southampton,
  SO17 1BJ, UK}
\footnotetext[3]{Department of Astrophysics, American Museum of
  Natural History, New York, NY 10024}
\footnotetext[4]{Space Telescope Science Institute, Baltimore, MD 21218}
\footnotetext[5]{South African Astronomical Observatory, PO Box 9,
  Observatory, 7935, South Africa} 
\footnotetext[6]{University of Helsinki, P.O. Box 14, SF-00014
  Helsinki, Finland}  
\footnotetext[7]{Astrophysics Group, School of Physics, Keele University,
  Staffordshire ST5 5BG, UK}  
\footnotetext[8]{Department of Astrophysics, Oxford University, Oxford
  OX1 3RH, UK} 
\begin{abstract}

We have used the Advanced Camera for Surveys on board the Hubble Space
Telescope to image the core of the globular cluster M\,15 in the
far-ultraviolet (FUV) waveband. Based on these observations, we identify
the FUV counterpart of the recently discovered low-mass X-ray binary
M\,15 X-2. Our time-resolved FUV photometry shows a modulation with
$0.062 \pm 0.004$ mag semi-amplitude and we clearly detect a period of
$22.5806 \pm 0.0002$ min. We have carried out extensive Monte Carlo
simulations which show that the signal is consistent with being
coherent over the entire observational time range of more than 3000
cycles. This strongly suggests that it represents the orbital period
of the binary system. M\,15 X-2 is FUV bright ($\rm{FUV} \simeq 17$ mag)
and is characterized by an extremely blue spectral energy distribution
($F_{\lambda} \propto \lambda^{-2.0}$). We also find evidence for an
excess of flux between 1500~\AA\ and 1600~\AA\ and probably between
1600~\AA\ and 2000~\AA\, which might be due to C\,IV 1550~\AA\  and
He\,II 1640~\AA\  emission lines. We also show that M15 X-2's X-ray
luminosity can be powered by accretion at the rate expected for
gravitational-wave-driven mass transfer at this binary period. The
observed FUV emission appears to be dominated by an irradiated
accretion disk around the neutron star primary, and the variability
can be explained by irradiation of the low-mass white dwarf donor if
the inclination of the system is $\approx 34^\circ$. We conclude that
all observational characteristics of M15 X-2 are consistent with it
being an ultracompact X-ray binary, only the third confirmed such
object in a globular cluster.   

\end{abstract}


\section{Introduction}

Ultracompact X-ray binaries (UCXBs) are very tight, interacting
binaries ($\approx 10^{10}$ cm orbit) with periods $\le 80$
min. They contain neutron stars (NSs) or black holes accreting from a
low-mass ($\le 0.1 {\rm M}_{\odot}$) degenerate companion. UCXBs are a
sub-class of the low-mass X-ray binaries (LMXBs). It has been known
for a long time that bright LMXBs are found to be overabundant in
globular clusters (GCs) compared to the Galactic field (Katz 1975,
Clark 1975). This leads to the suggestion that LMXBs, and 
consequently UCXBs, are preferentially formed in the dense environment
of the GC cores through various stellar interactions, such as
direct collisions between a NS and a red giant (Verbunt 1987, Davies
et al.~1992, Ivanova et al.~2005), tidal capture of a main sequence
star by a NS (Bailyn \& Grindlay 1987), or exchange interactions
between NSs and primordial binaries (Rasio et al.~2000). 
It has been suggested that most of the 13 bright LMXBs in GCs might be
UCXBs (Bildsten \& Deloye 2004, Ivanova et al.\ 2005). However, only
two are confirmed in GCs to date. These are 4U\,1820-30 in NGC\,6624 
($P_{\rm{orb}} = 11.4$ min, Stella et al.~1987)
and 4U\,1850-087 in NGC\,6712 ($P_{\rm{orb}} = 20.6$ min, Homer et
al.~1996). Two other GC sources have been suggested as particularly
strong UCXB candidates (Homer 2003).
One of these candidates is CXO\,J212958.1+121002
in M\,15, also known as M\,15 X-2 (White \& Angelini 2001). 

M\,15 is the only galactic GC known to harbour {\it two} bright
LMXBs. A single source was detected in early X-ray studies,
4U\,2127+119, and that was identified  with the optical counterpart
AC\,211 (Auri{\`e}re et al. 1984, Charles et al. 1986). 
{\it Chandra} observations later on resolved 4U\,2127+119 into {\it
  two} X-ray sources (White \& Angelini 2001). One of these was the
previously known LMXB AC\,211. The second source, CXO\,J212958.1+121002 
or M\,15 X-2, is actually 2.5 times brighter than AC\,211 in X-rays. 
Based on {\it Hubble Space Telescope (HST)} data from Guhathakurta et
al.\ (1996), White \& Angelini (2001) identified a blue $U = 18.6$ mag
star as the optical counterpart to the second source (star 590 in
De\,Marchi \& Paresce 1994). However, the orbital period of
M\,15 X-2 has so far not been determined. 

Here, we present FUV data of M\,15 X-2 taken with the {\it HST} that allow
us to classify the source as an UCXB. In 
Sect.~\ref{data} we describe the data and their reduction. We present
the analysis of the photometry and determine the period of M\,15 X-2
in Sect.~\ref{analysis}. We summarize our results and conclusions in
Sec.~\ref{summary}. 

\section{Observations and data reduction}
\label{data}

M\,15 was observed with the Advanced Camera for Surveys (ACS) on board
the {\it HST} in  September and October 2003, and October to December
2004. Images were taken using the far-ultraviolet (FUV) filters F140LP,
F150LP, and F165LP in the Solar Blind Channel (SBC), and the near-UV (NUV)
F220W filter in the High Resolution Channel (HRC). 

For our variability study (see Sect.~\ref{analysis}) only relative
magnitudes are needed. These were derived directly from the
individual flatfielded images. Aperture photometry was carried out
using {\tt daophot} (Stetson 1991) running under {\tt IRAF}
\footnote{{\tt IRAF} (Image 
  Reduction and Analysis Facility) is distributed by the National
  Astronomy and Optical Observatories, which are operated by AURA,
  Inc., under cooperative agreement with the National Science
  Foundation.} 
using an aperture radius of 4 pixels and a sky annulus of 50 to 60
pixels. For this purpose, we only used the 90 exposures taken in
SBC/F140LP, since these data provide the longest  time coverage
(from October 14 to December 5, 2004). Eighty of these images had
exposure times of 300 sec, four were exposed for 140 sec and six for
40 sec. One orbit of observation yielded 9 data points.   

In order to determine the time-averaged spectral energy distribution
(SED) of our counterpart, we also carried out {\it absolute}
photometry on combined and geometrically corrected images for each
filter. These master images were created using {\tt multidrizzle}
running under {\tt PyRAF}. For our aperture photometry, we used an
aperture radius of 5 pixels for all SBC FUV data, a smaller radius of
4 pixels in the HRC NUV data, and a sky annulus of 5 to 7 pixels. The
smaller aperture in the NUV was chosen to avoid the effects of severe
crowding. In the FUV, aperture corrections were determined via curves
of growth constructed from isolated stars in our master images. For
the NUV data, we used the encircled energy fractions published by
Sirianni et al. (2005). 
In order to derive a reliable SED, we ideally want mean fluxes in
non-overlapping wavelength windows. However, the ACS/SBC/F140LP
bandpass fully includes F150LP, and F150LP in turn includes F165LP. We
therefore created two artificial narrow-band filters that are defined
as the differences between the actual filters, i.e.\ we
define  F140N = F150LP - F140LP and F150N = F150LP - F165LP. Thus the
count rate of a source in F140N, for example, is simply obtained by
subtracting its count rate in F150LP from that in
F140LP. Fig.~\ref{spect} (top panel) shows the resulting
throughput curves for the artificial filters. As can be seen, they  barely
overlap and are thus ideal for characterizing the UV SED of our source.   
In order to convert count rates into STMAGs, we used the {\tt synphot}
package running under {\tt IRAF}. Full details on our analysis
procedure will be provided in a forthcoming publication that will
present our entire FUV/NUV data set for M\,15. 

\section{Analysis and Discussion}
\label{analysis}

\subsection{Source Identification}

\begin{figure}
\epsscale{1.09}
\includegraphics[scale=0.565]{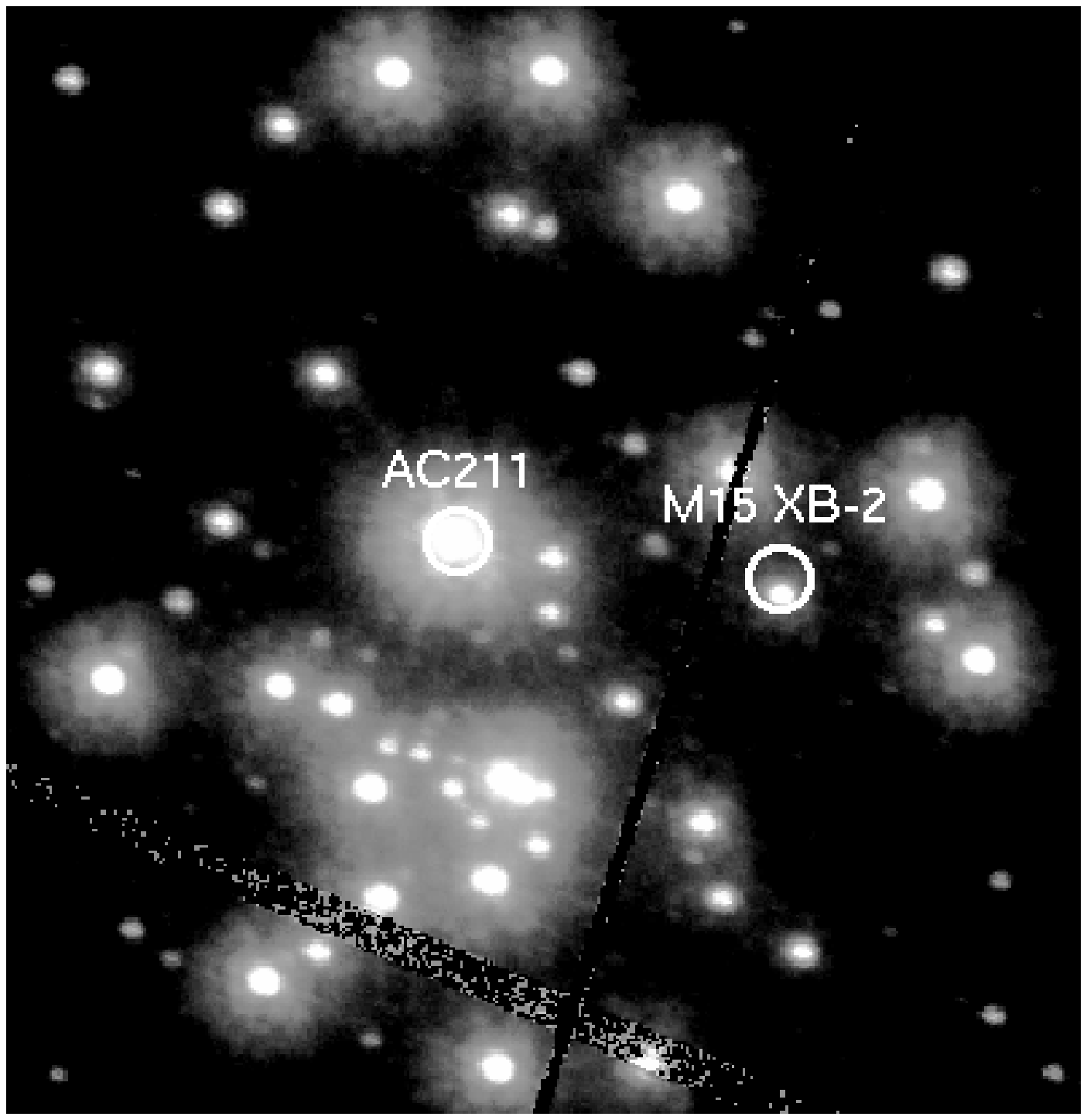}
\includegraphics[scale=0.5]{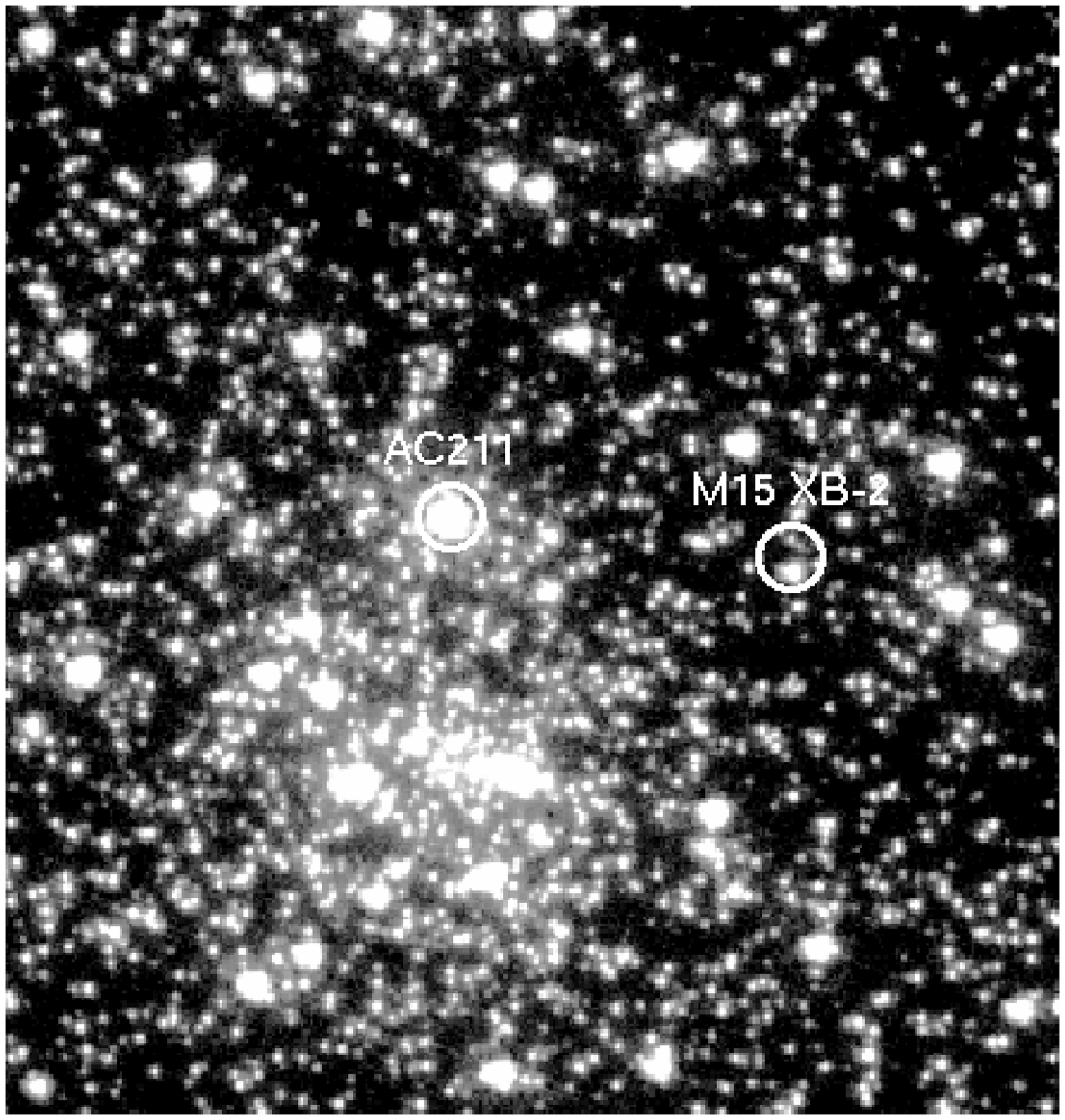}
\caption{Left: FUV image (taken in SBC/F140LP) of M\,15 X-2. North
  is up and east to the left. The field-of-view of this close-up is
  $\approx 8\farcs9 \times 8\farcs9$. The {\it Chandra} positions for
  AC\,211 and M\,15 X-2 are indicated. 
  Right: The same but for the NUV. Note
  the severe crowding in this image. Both the LMXB AC\,211 and the
  UCXB M\,15 X-2 stand out as bright objects in both
  images. \label{XB2}} 
\end{figure}

We identified the FUV counterparts to AC\,211 and to M\,15 X-2 on the
F140LP images, using the Guhathakurta et al.~(1996) {\it HST} images and the
positions provided by White \& Angelini (2001) as a reference
guide. Fig.~\ref{XB2} shows a close-up of the FUV (SBC/F140LP) and NUV
(HRC/F220W) master images, centred on AC\,211 and M\,15 X-2. The
{\it Chandra} positions for these sources are also indicated, after shifting
them by $\approx 1\farcs2$~south in order to optimally align the X-ray
and FUV positions of AC\,211. The centre of the FUV counterpart to
M\,15 X-2 is slightly offset ($0\farcs165$ south) of the {\it
  Chandra} position, but still well within the internal $0\farcs25$
{\it Chandra} error radius for this source (White \& Angelini
2001). Our offset is consistent with the $0\farcs13$ offset that White
\& Angelini (2001) found between their {\it Chandra} positions and the
optical counterpart from De\,Marchi \& Paresce (1994). Both AC\,211
and M\,15 X-2 are clearly detected as strong FUV and NUV sources in
our ACS images.    

\subsection{Time-Series Analysis}

\begin{figure}
\epsscale{0.98}
\includegraphics[angle=270,scale=0.63]{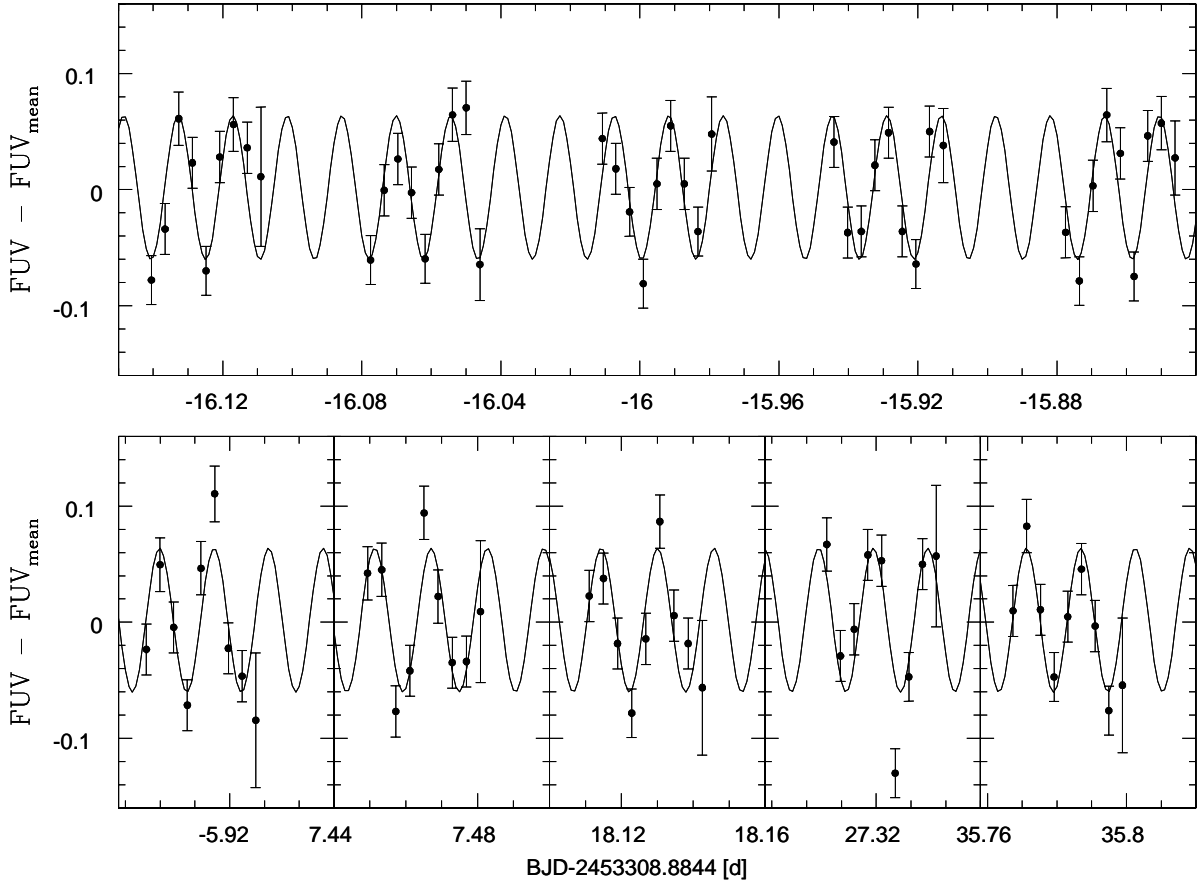}
\caption{Lightcurves of M\,15\,X-2 (mean subtracted photometric data) with
  overplotted sine wave of 22.5806 min period. All six observing
  epochs are plotted on the same scale, note that the top panel
  displays the first observing epoch which lasted for 5 orbits, the
  bottom panel displays the following 5 epochs which lasted for one
  orbit each. \label{lightcurve}}
\end{figure}

\begin{figure}
\includegraphics[angle=270,scale=0.45]{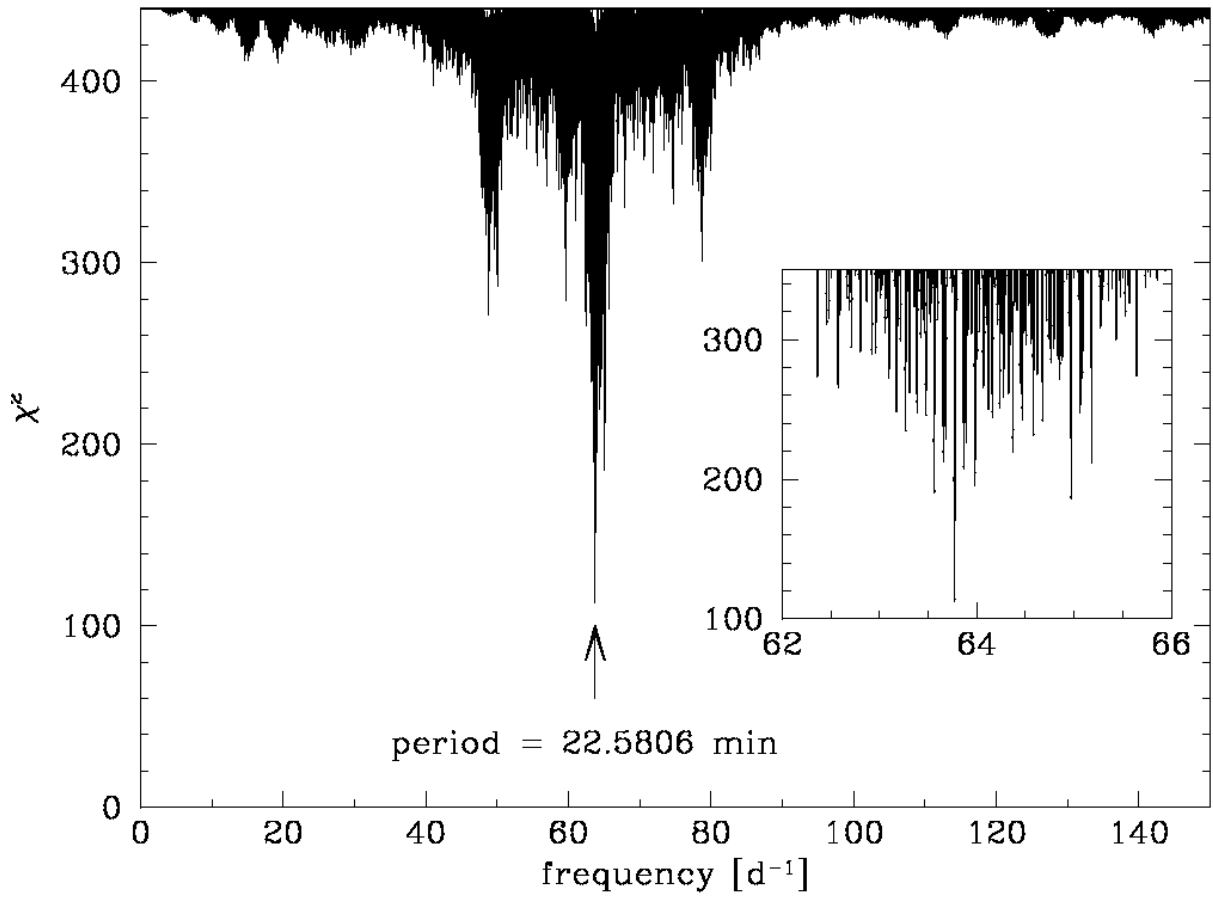}
\caption{$\chi^{2}$ vs. frequency periodogram of M\,15\,X-2, the
  strong peak corresponds to the period of 22.5806 min. The inset
  figure shows a zoom on the peak frequencies at 63.77 d$^{-1}$. See
  the text for details. \label{power}}      
\end{figure}

Fig.~\ref{lightcurve} shows the mean-subtracted light curve for all
observing epochs. Low-amplitude variations with a peak-to-peak
amplitude $> 0.1$ mag can be seen. We searched for a periodic signal
by carrying out $\chi^{2}$ fits for a grid of trial frequencies. 
The resulting periodogram is shown in
Fig.~\ref{power}. The best fit yields a $\chi_{\nu}^{2}=1.31$ and
suggests a period of $22.5806$ 
min. In order to test the coherence of this period, we carried out
Monte Carlo simulations. Briefly, we created 10000 fake data sets with
the same time sampling, periodic signal and noise
characteristics as our real data, but with a random phase offset
assigned to each of the six observing epochs. Thus the phase was fully
coherent within each epoch, but fully randomized between them. 
We then again carried out a sequence of $\chi^{2}$\ fits for each data
set. Next, we fixed the period at the best global value and fitted
each epoch separately with both phase and amplitude as free
parameters. We then defined a {\it phase coherence index} (PCI, see
e.g.\ Haswell et al.\ 1997), which in our case is simply the
$\chi^{2}$ of the phase estimates for the individual epochs, with the
phase of the global fit as the reference value. {\it We find that only
  0.9\% of the fake data with randomized phases have a PCI as good as
  the real data.} We can therefore reject the null hypothesis that the
periodic signal loses coherence completely over time-scales comparable
to our inter-epoch spacing. The latter is $\approx 11$ days,
corresponding to $\approx 700$ cycles. We can view this as a
constraint on the period derivative, i.e. $\dot{P}$ must be small
enough so that less than one cycle is lost over $N \simeq 700$ cycles,
i.e.\ $\dot{P} \lesssim N^{-1}$. The quality factor of the 22.58~min
signal must therefore be $Q = \dot{P}^{-1} \gtrsim 700$. By contrast,
mHz QPOs tend to have $Q \approx 10$ (e.g. Chakrabarty et al.\ 2001,
Boroson et al.\ 2000). As a further check, we carried out Monte Carlo
simulations in which the input sinusoids were coherent, i.e. the phase
was fixed at the same value for each epoch. This produced a PCI
distribution that was consistent with the PCI of the real data
set. Thus the 22.58~min signal is consistent with being fully coherent
over the entire $\simeq 3300$\ cycles spanned by our observations. All
of these tests support the orbital nature of this signal.   

We also used our Monte Carlo simulations to estimate the statistical
error on the parameters of the observed signal. For this purpose, we
again used coherent input signals and conservatively used error bars
scaled so as to yield $\chi_{\nu}^{2} =1$ for the fit to the real
data. We then used the standard deviation of the periods and amplitudes found
for the fake data sets to estimate the errors on the measured
parameters. The final results were $P = 22.5806 \pm 0.0002$ min for
the orbital period and $a = 0.062 \pm 0.004$ mag for the
semi-amplitude. 
This yields an ephemeris for the time of maximum light
\begin{equation}
T_{max}\rm{(BJD)} = 2453308.88743(16) + 0.01568096(13) \times E,
\end{equation}
where the numbers in brackets give the errors on the last two digits. 
  
A sine wave with M\,15\,X-2's period, amplitude and fixed phase is
overplotted on the lightcurves in Fig.~\ref{lightcurve}. 
This visually confirms that there is no sign of a loss of coherence
over the entire observational time span of 3312 cycles. We conclude
that the periodic signal is almost certainly an orbital modulation. 
\footnote{We note that the average time resolution of our data is
  $\approx 5.5$ min which corresponds to a Nyquist frequency of
  $\nu_{Ny} \approx 130 \rm{d}^-1$. A period of $\approx 7.4$ min
  above $\nu_{Ny}$ could be reflected to yield our observed signal of
  22.6 min. However, such short orbital periods are extremely unlikely
  (see e.g. Deloye \& Bildsten 2003, Homer 2003).}

\subsection{Continuum Spectral Energy Distribution}

The SED of M\,15 X-2 is shown in Fig.~\ref{spect} (top panel). Each
point is plotted at the average wavelength of the corresponding
filter. As expected for compact, interacting binaries, the SED of M\,15 X-2
rises towards the blue. However, it is also worth noting that there
seems to be an excess of flux in F150N, i.e.\ around 1550~\AA. This
excess flux can be caused by additional C\,IV $\lambda = 1550$~\AA\
and/or He\,II $\lambda = 1640$~\AA\ line emission. We therefore fit a
power law to the two bracketing data points only. The best fit is found for a
power-law index $-2.0 \pm 0.2$. This power-law spectrum is
overplotted in Fig.~\ref{spect}. The depression at $\approx 2200$ \AA\
is due to a well known reddening feature there.  

\subsection{Evidence for Line Emission}
\label{ev_line}

The excess flux in F150N seems to indicate line emission due to C\,IV
at~$\lambda = 1550$~\AA\ and/or He\,II $\lambda = 1640$~\AA\ (the
latter would also contribute to F165LP). However, such a peak might
also be caused simply by a turnover in an otherwise smooth 
continuum. We have therefore carried out synthetic photometry 
for blackbodies (BBs) with temperatures
$100000 \ge T_{eff} \ge 10000$. Fig.~\ref{spect} (bottom panel) shows
the resulting BB sequence in the F140N-F150N vs F150N-F165LP
colour-colour diagram (CC diagram). Note that we reddened all
synthetic photometry by M\,15's $E_{(B-V)}=0.1$ mag (Harris 1996). As
expected, the BBs are located on a sequence going from blue (for hot
sources) to red colours (for cool sources). The cross on the sequence marks
$T_{eff} = 20200$ K, which is the temperature of a BB peaking at
$\lambda = 1550$~\AA, in the centre of the F150N filter.
The observed location of M\,15 X-2 in the CC diagram is
also marked and is clearly distinct from the blackbody sequence. The
reason is simply that the F150N filter is much narrower than the peaks
of the BB distributions. Thus, the latter cannot cause a strong excess
in this filter alone. We caution that true stellar spectra {\it can}
have turnovers more sharp than suggested from a comparison with BB
spectra, as the example of AC\,211 shows (Downes et al.\ 1996).   

The location of a power-law $F_{\lambda} \propto \lambda^{-2.0}$
spectrum is also marked in the CC diagram.  
In order to check how strong a line might be needed to account for the
observed flux excess, we have also carried out synthetic photometry of
power law spectra (with index -2.0) {\it and} an emission line at C\,IV 
$\lambda = 1550$~\AA\ with equivalent widths (EW) of 10~\AA\ to
50~\AA. As can be seen, M\,15 X-2 is located close to the power-law +
C\,IV 1550~\AA\ sequence, but slightly above, which suggests additional He\,II
emission. We then carried out synthetic photometry of power law
spectra and C\,IV $\lambda = 1550$~\AA\ ($\rm{EW} \simeq 30$ \AA) {\it
  and} He\,II $\lambda = 1640$~\AA\ emission line with EWs of 10~\AA\
to 60~\AA. We conclude that the SED of M\,15 X-2 can be described by
a power-law $F_{\lambda} \propto \lambda^{-2.0}$ with an additional
C\,IV 1550~\AA\ and He\,II 1640~\AA\ emission line with $\rm{EW}
\simeq 30$~\AA\ each. However, given that there are three free
parameters in this model, a perfect match to just three colours is of
course guaranteed. Spectroscopy will be needed to confirm the spectral
shape and the existence of line emission.

\section{Discussion}
\label{summary}

Knowledge of the orbital period of M\,15 X-2 allows us to derive a
more defined picture of this ultracompact system. Eggleton (1983)
showed that for small mass ratios $0.01 \le q = M_{2}/M_{1} \le 1$ the
mean density $\rho$ of the Roche lobe-filling companion becomes a
function of $P_{orb}$ mainly, i.e.\ $P_{orb} \times \rho^{1/2} \simeq
0.438$. For an orbital period of 22.6 min, this gives $\rho = 786~
\rm{g}~ \rm{cm}^{-3}$ which is consistent with the mean density of a
low-mass white dwarf (WD). We can then use the mass-radius relationships
published by Deloye \& Bildsten (2003, their Fig.~4) to constrain the
minimum mass of the donor star to $0.02~\rm{M}_\odot \le M_{2,min}
\le 0.03~\rm{M}_\odot$ and its minimum radius to $0.03~\rm{R}_\odot
\le r_{2,min} \le 0.04~\rm{R}_\odot$, depending on composition. These
lower limits correspond to low-temperature donors. Using Kepler's 3rd
law and assuming a NS-dominated system mass near $1.4~\rm{M}_\odot$,
we estimate the binary separation to $\approx 2.1 \times 10^{10}$ cm.     

\begin{figure}
\plotone{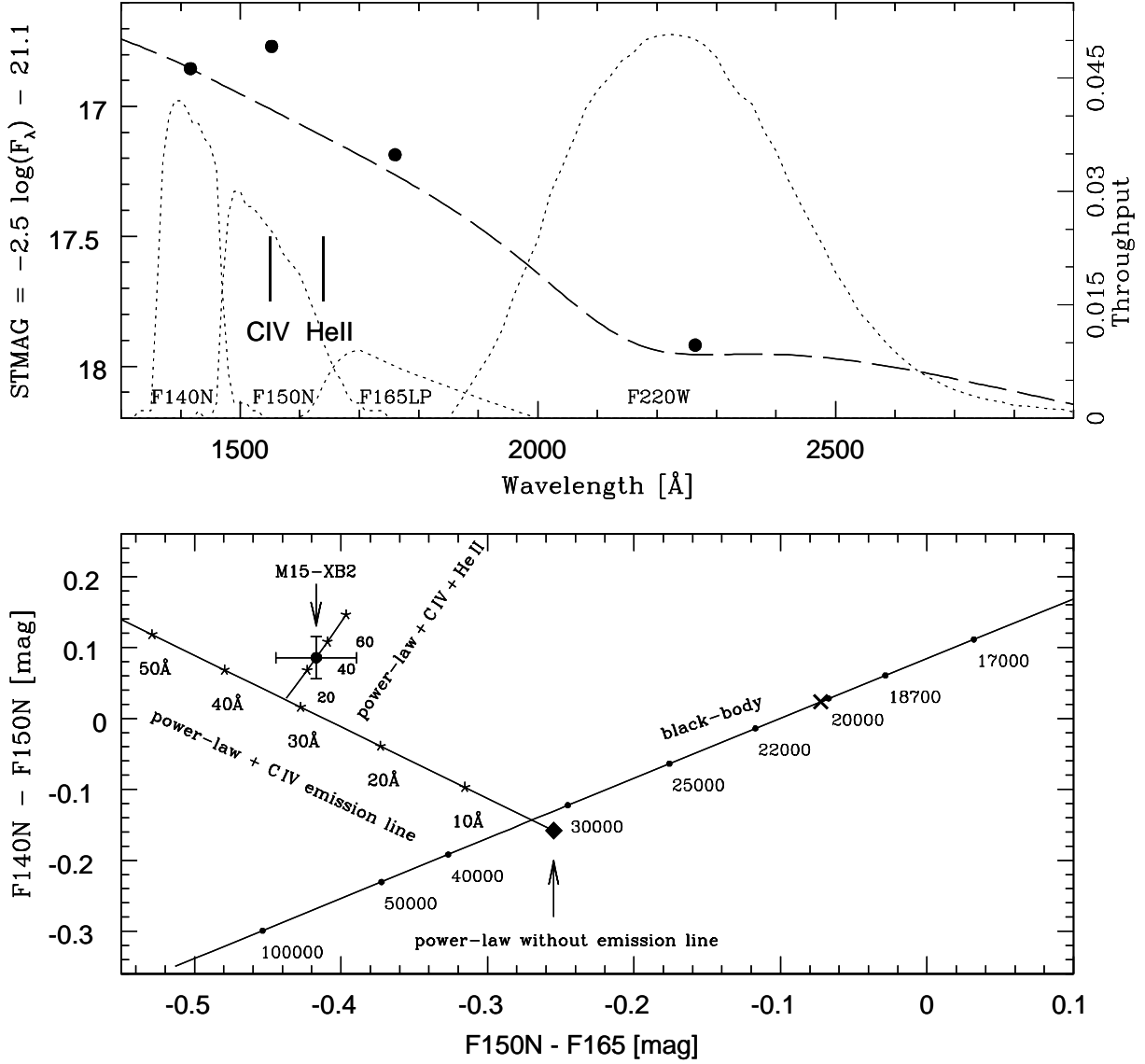}
\caption{Top panel: Spectral energy distribution of M\,15\,X-2. The
  dashed line corresponds to a power-law $F_{\lambda} \propto
  \lambda^{-2.0}$. Filter transmission curves 
  are overplotted as dotted lines. Bottom panel: Colour-Colour diagram for 
  BBs of different $T_{eff}$ (dots). The cross on the BB sequence
  marks $T_{eff} = 20200$ K, which is the temperature of a BB peaking
  at $\lambda = 1550$ \AA. The filled diamond denotes a power-law with
  index -2.0. The stars denote power law spectra (with index -2.0)
  with additional line emission with equivalent widths (EW) of 10\AA\
  to 60 \AA. M\,15\,X-2 is also marked. 
  See the text for details. \label{spect}} 
\end{figure}

A blackbody of $T_{eff} \approx 32000$ K has a blue spectral slope
most similar to the one we fitted to M\,15 X-2 (see
Fig.~\ref{spect}). Placed at M\,15's distance, such a blackbody would
have a radius of $r_{bb} \approx 6.5 \times 10^{9}$ cm or $0.1\
\rm{R}_\odot$ if it is to have the same flux that we measured for
M\,15 X-2. This is larger than expected for the radius $r_{2}$ of the
degenerate companion, but comparable to the circularization radius of 
$r_{circ} \approx 0.2\ \rm{R}_\odot$ of the accretion disk.
We therefore conclude that the FUV light is coming from the accretion
disk rather than from the WD donor. 
This is consistent with the UCXB model of Arons \& King (1993)
in which the orbital modulation is then caused by the irradiation
of the WD donor. Using their Eq.~15 we then estimate the
inclination angle of the system $i \approx 34^{\circ}$. This face-on
inclination is consistent with the absence of eclipses. 
We note that no modulation can be seen in X-rays (Hannikainen et
al. 2005).
  
M\,15 X-2's X-ray luminosity was found to be $L_{X} \approx 1.4 \times
10^{36}\ \rm{erg}\ \rm{s}^{-1}$ (White \& Angelini 2001, Hannikainen
et al.\ 2005). Assuming a 10 km radius and a mass of $1.4\
\rm{M}_\odot$ for the NS, this requires $\dot{M} > L_X R_\star / 
G M_\star \approx 10^{-10}\ \rm{M}_\odot\ \rm{yr}^{-1}$. This can be
compared to the accretion rate expected from conservative mass
transfer driven by angular momentum loss via gravitational radiation
in an UCXB  
\begin{equation}
\dot{M_{gr}} = 1.27 \times 10^{-8} \times q^{2} \times M_\star^{8/3}
  \times P_{orb}^{-8/3}(\rm{h}) \times (1+q)^{-1/3} \times (5/6 + n/2
  -q)^{-1}. \label{mdot}
\end{equation}
Taking our minimum donor mass and corresponding mass-radius index
$n \simeq -0.1$ (e.g. Deloye \& Bildsten 2003), we 
derive a lower limit of $\dot{M_{gr}} \approx 4 \times 10^{-10}\ \rm{M}_\odot
yr^{-1}$. This suggests that the observed X-ray emission can be
powered by gravitational radiation-driven mass transfer. We conclude
that M\,15 X-2 can be classified as an UCXB, only the third confirmed
such system in a GC. Our results are consistent with the idea that
indeed many GC LMXBs are UCXBs.

\acknowledgments

We thank Tom Maccarone, Tom Marsh, Geoff Daniell and Chris Deloye for
valuable discussions. This work was supported by NASA through grant
GO-9792 from the Space Telescope Science Institute, which is operated
by AURA, Inc., under NASA contract NAS5-26555.

\end{document}